
%
\input harvmac
\lref\Affrev{For a review, see I. Affleck, ``Conformal Field Theory Approach to
Quantum Impurity Problems'', UBCTP-93-25, cond-mat/9311054.}
\lref\AFL{N. Andrei, K. Furuya, and J. Lowenstein, Rev. Mod. Phys. 55
(1983) 331; \hfill\break
A.M. Tsvelick and P.B. Wiegmann, Adv. Phys. 32 (1983) 453.}
\lref\RS{N. Reshetikhin and H. Saleur, ``Lattice Regularization of Massive and
Massless Field Theories'', USC-93-020, hep-th/9309135.}
\lref\KR{A.N. Kirillov and N. Reshetikhin, J. Phys. A20 (1987) 1565, 1587.}
\lref\TS{M. Takahashi and M. Suzuki, Prog. Th. Phys. 48 (1972) 2187.}
\lref\chered{I. Cherednik, Theor. Math. Phys. 61 (1984) 977.}
\lref\GZ{S. Ghoshal and A.B. Zamolodchikov, ``Boundary State and Boundary $S$
Matrix in Two-Dimensional Integrable Field Theory'', RU-93-20, hep-th/9306002.}
\lref\fki{A. Fring and R. K\"oberle, ``Factorized Scattering in the Presence of
Reflecting Boundaries'', USP-IFQSC/TH/93-06, hep-th/9304141.}
\lref\fkii{A. Fring and R. K\"oberle, ``Affine Toda Field Theory in the
Presence of  Reflecting Boundaries'', USP-IFQSC/TH/93-12, hep-th/9309142.}
\lref\ghosi{S. Ghoshal, ``Bound State Boundary $S$ Matrix of the Sine-Gordon
Model'',  RU-93-51, hep-th/9310188.}
\lref\ghosii{S. Ghoshal, ``Boundary $S$ Matrix of the $O(n)$ Symmetric
Nonlinear Sigma Model'' RU-94-02, hep-th/9401008.}
\lref\sasaki{R. Sasaki, ``Reflection Bootstrap Equations for Toda Field
Theory'', 
 hep-th/9311027.}
\lref\ABBBQ{F. Alcaraz, M. Barber, M. Batchelor, R. Baxter and G. Quispel, J.
Phys. A20 (1987) 6397.}
\lref\skly{E.K. Sklyanin, J. Phys. A21 (1988) 2375.}
\lref\ZandZ{A.B. Zamolodchikov and Al.B. Zamolodchikov, Ann.  Phys.
120 (1980) 253.}
\lref\AL{I. Affleck and A. Ludwig, Phys. Rev. Lett. 67 (1991) 161.}
\lref\CarVer{J. Cardy, Nucl. Phys. B324 (1989) 581.}
\lref\EA{S. Eggert and I. Affleck, Phys. Rev. B46 (1992) 10866.}
\lref\pkondo{P. Fendley, Phys. Rev. Lett. 71 (1993) 2485, cond-mat/9304031.}
\lref\FS{P. Fendley and H. Saleur, Nucl. Phys. B388 (1992) 609,
hep-th/9204094.}
\lref\DL{C. Destri and J.H. Lowenstein, Nucl. Phys. B205 (1982) 369.}
\lref\DNW{G.I. Dzapardize,  A.A. Nersesyan and P.B. Wiegmann, Phys. Scr. 27
(1983) 5.}
\lref\DV{C. Destri and H. de Vega, J. Phys. A22 (1989) 1329.}
\lref\boundrefs{L. Mezincescu and R.I. Nepomechie, Int. J. Mod. Phys. A6 (1991)
 5231; Int. J. Mod. Phys. A7 (1992) 565; H.J. de Vega and A. Gonzalez-Ruiz,
 preprint LPTHE-93/38.}
\lref\nonstrings{F. Woynarovich, J. Phys. A15 (1982) 2985;\hfill\break
O. Babelon, H.J. de Vega and C.M. Viallet, Nucl. Phys. B220
 (1983) 13.}
\lref\EKS{F. Essler, V. Korepin and K. Schoutens, J. Phys A25 (1992) 4115.}

\def\t{\theta}
\def\g{\gamma}
\def\<{\langle}
\def\>{\rangle}
\noblackbox
\Title{\vbox{\baselineskip12pt
\hbox{USC-94-001}\hbox{hep-th/9402045}}}
{\vbox{\centerline{Deriving boundary $S$ matrices
}}}

\centerline{P. Fendley$^\dagger$ and H. Saleur$^{\spadesuit *}$}
\vskip2pt
\centerline{$^\dagger$Department of Physics, University of Southern California}
\centerline{Los Angeles CA 90089}
\vskip2pt
\centerline{$^\spadesuit$Department of Physics and Department of Mathematics}
\centerline{University of Southern California}
\centerline{Los Angeles CA 90089}
\vskip.3in
We show how to derive  exact  boundary $S$ matrices  for  integrable quantum
field theories in 1+1 dimensions using  lattice regularization. We
do this calculation explicitly for the sine-Gordon model with fixed boundary
conditions using the Bethe ansatz for an XXZ-type spin chain in a
boundary magnetic field. Our results agree with recent conjectures of Ghoshal
and Zamolodchikov, and  indicate  that the only solutions to the Bethe
equations which contribute to the scaling limit are the standard strings.

\bigskip
\bigskip\bigskip
\noindent $^*$ Packard Fellow
\Date{2/94}

\noindent{\bf 1. Introduction}
\smallskip

Many interesting statistical-mechanical systems can be described by
1+1-dimensional models with boundaries. These include the Kondo problem and the
Anderson model, tunneling in quantum wires, dissipative quantum mechanics, and
the Callan-Rubakov effect. At and near a critical point, they can be studied
using boundary conformal field theory \Affrev. Moreover, many of these models
are integrable and many properties can be derived exactly even away from
critical points \AFL.

There are several approaches to integrable models, with and without a boundary.
One is explicit diagonalization of the (continuum or suitably
discretized) Hamiltonian using the Bethe ansatz. Another bypasses this
diagonalization by using maximally the constraints resulting from
integrability: after making some guesses for the particle content one can find
physical quantities like the exact $S$ matrix for
particles
scattering among themselves and off the boundary. A crucial ingredient in both
cases is the Yang-Baxter equation, including the boundary part  \chered. In
the second approach this
equation only allows one to find ratios of $S$ matrix
elements and not their overall prefactor. In the bulk such prefactors are
controlled by unitarity and crossing symmetry. For the boundary,
 constraints of the same type were found in \refs{\GZ,\fki}, enabling the
determination of $S$ matrix elements up to the usual CDD-type ambiguities. This
was done for the Ising model in a boundary magnetic field \GZ, the sine-Gordon
model with a boundary potential \refs{\GZ,\ghosi}, affine Toda field theories
\refs{\fkii,\sasaki} and the $O(n)$ sigma models with free or fixed boundary
conditions \ghosii.

The purpose of this paper is to use the first approach to confirm the second.
We extract boundary $S$ matrices from lattice regularizations of integrable
quantum field theories, generalizing a well-known method in the bulk
\refs{\DL,\DNW,\KR}. Our explicit example is
the inhomogeneous XXZ model in a boundary
magnetic field, which is expected to provide an integrable regularization
of the sine-Gordon model with fixed boundary
conditions.  Our results confirm results of \GZ, in particular their
cross-unitarity relation, and remove the CDD ambiguity. The interest of the
computation lies mainly in the fact that boundary effects are subleading, and a
number of subtleties arise in their analysis.

We derive the boundary $S$ matrix elements in the usual  manner. We
start with the explicit diagonalization of the appropriate lattice model
with integrable boundary conditions. This
results in the ``bare'' Bethe ansatz equations, which relate the level
densities of the bare particles to the actual filled densities of these
particles. Using these equations, we find the ground state of the system,
fill the Fermi sea, and identify physical excitations.  This results in
the ``physical'' Bethe ansatz equations,
which involve the densities of the actual quasiparticles of the system.
 On the other hand, physical Bethe-ansatz equations can also be derived
 directly from the conjectured $S$
matrix, without reference to a lattice. We perform such a computation
for the $S$ matrix of \refs{\GZ,\ghosi}. We compare the two results, and find
that they are indeed the same.

The lattice model of interest is the inhomogeneous 6-vertex model
\refs{\DV,\RS}\ on an open strip, with integrable boundary conditions as in
\skly. We take identical boundary conditions on both sides of the system. In
the hamiltonian and homogeneous limit, this model reduces to
the XXZ model with boundary magnetic fields $h$ \ABBBQ
\eqn\ham{{\cal H}= \epsilon{\gamma\over 2 \pi\sin\gamma}
\left[\sum_{i=1}^{N-1} \left(
\sigma^x_i \sigma^x_{i+1} + \sigma^y_i \sigma^y_{i+1}- \cos \gamma\sigma^z_i
\sigma^z_{i+1}\right) +h\sigma_1^z+h\sigma_N^z\right].}
In the inhomogeneous 6-vertex model, one gives an alternating  imaginary part
$\pm i\Lambda$ to the spectral parameter on alternating vertices
\refs{\DV,\RS}. The scaling limit is given by taking
$\Lambda\rightarrow\infty$, $N\rightarrow\infty$, and the lattice spacing
$\Delta\rightarrow 0$, such that $L\equiv N\Delta$ remains finite. In the bulk,
this provides a regularization of the sine-Gordon model with
Lagrangian
\eqn\sg{L_{\hbox{SG}}
=\int_0^L dx\ \left[ \half (\del\phi)^2 + \mu^2 \cos \beta_{SG}
\phi\right]}
where the mass $\mu\propto {1\over\Delta}\exp(-{\rm const}\Lambda)$, and
$\beta_{SG}^2=8(\pi-\gamma)$ for the antiferromagnetic case ($\epsilon=-1$),
while $\beta_{SG}^2=8\gamma$ for the ferromagnetic one ($\epsilon=1$). The
scaling limit of the gapless hamiltonian \ham\ is given
by the ultraviolet limit of \sg.

Free boundary conditions in the lattice model
correspond to fixed ones in the sine-Gordon model.
This can easily be seen by recalling that $\phi$ is dual to the arrows of the
vertex model; since in the approach of \refs{\DV,\RS} the time direction is
across the diagonal of a vertex, the field at the boundary is necessarily
constant. Thus we have
$\phi(0)=\phi(L)=\phi_0$, where $\phi_0$ depends on the boundary magnetic field
$h$ (at $h=0$ we have $\phi_0=0$). These boundary conditions preserve the
topological $U(1)$ symmetry giving conservation of soliton number, but they
break the $Z_2$ charge conjugation symmetry relating soliton and antisoliton
(unless $h=0$).

Because we are mainly interested in checking the physical $S$ matrices
and paving the way towards a thermodynamic analysis, we shall discuss the
lattice model
and the underlying inverse scattering problem in the presence of boundaries as
little as possible.
We refer the reader to \refs{\skly,\boundrefs} for details. We shall simply use
the resulting bare Bethe ansatz equations without
deriving them.

\bigskip
\noindent{\bf 2. The bare Bethe ansatz equations}
\smallskip

The wave function of the inhomogeneous six-vertex model can be expressed in
terms of a set of ``roots''  $\alpha_j$, where $j=1\dots n$. They must be
solutions of the set of equations \refs{\ABBBQ,\skly}
\eqn\foral{\eqalign{N\left[f(\alpha_j+\Lambda,\gamma)+f
(\alpha_j-\Lambda,\gamma)\right]+
2f(\alpha_j,\gamma H)=\cr
2\pi l_j + \sum_{m=1,m\ne
j}^n \left[f(\alpha_j-\alpha_m,2\gamma) +
f(\alpha_j+\alpha_m,2\gamma)\right],\cr}}
where $l_j$ is an integer. The function $f$ is defined as
$$f(a,b)=-i \ln\left({\sinh(ib-a)/2 \over \sinh(ib+a)/2}\right)$$
and
\eqn\forH{H\equiv{1\over\gamma} f(i\gamma,-i\ln(h+\cos\gamma)).}
For $h=0$, $\gamma H=\pi-\gamma$. By construction
 of the Bethe-ansatz wave function, $\alpha_j>0$.
Even though there is a solution of
 \foral\ with one vanishing root for any $N$ and $n$,
we emphasize that $\alpha_j=0$ is not allowed  because the wave function
vanishes identically in this case.

The solutions of these equations are quite intricate for arbitrary $\gamma$
\TS. For simplicity, we restrict to
the case $\gamma=\pi/t$  where $t$ is an integer, and consider both choices
$\epsilon=\pm 1$.
In the sine-Gordon model, these fall in the attractive and repulsive regimes
respectively. We make the standard assumption that all the solutions of
interest are collections of  ``$k$-strings'' for $k=1,2\dots t-1$ and
antistrings $a$
\TS. A $k$-string is a group of $\alpha_j$ in the pattern
$\alpha^{(k)}-i\pi(k-1),  \alpha^{(k)}-i\pi(k-3), \dots, \alpha^{(k)}+
i\pi(k-1)$ where $\alpha^{(k)}$ is real. The antistring has
$\alpha_j=\alpha^{(a)}+i\pi$, where  $\alpha^{(a)}$ is real.
 We shall comment on the validity of this assumption (which
is more crucial when one considers subleading effects) in the conclusion.

The thermodynamic  limit is obtained by sending $N\to\infty$. In this case,
we can
define densities of the different kinds of solutions.
The number of allowed solutions of
\foral\ of type $k$ in the interval $(\alpha,\alpha+d\alpha)$ is
$(\rho_k(\alpha) + \rho_k^h(\alpha)) 2N d\alpha$, where $\rho_k$ is the density
of ``filled'' solutions
(those which appear in the sum in the right-hand-side of \foral\ ) and
$\rho_k^h$ is the density of ``holes'' (unfilled solutions). The densities
$\rho_a$ and $\rho_a^{h}$ are defined likewise for the antistring. The ``bare''
Bethe ansatz equations follow from taking the derivative of $\foral$. For
$\gamma=\pi/t$ they can be written in the form:
\eqn\bba{\eqalign{2\pi (\rho_k + \rho_k^h)&=  {1\over 2}\left[a_k
(\alpha+\Lambda)+a_k(\alpha-\Lambda)\right] -
\phi_{k,t-1}*\rho_{a} +\sum_{l=1}^{t-1}
\phi_{kl}*\rho_l +{1\over 2N}u_k\cr
2\pi(\rho_{a} +\rho_a^h)&=2\pi(\rho_{t-1}+ \rho_{t-1}^h) +
{1\over 2N}(u_a-u_{t-1})\cr}}
where $*$ denotes convolution:
$$ f*g(\alpha)\equiv \int_{-\infty}^{\infty} d\alpha' f(\alpha-\alpha')
g(\alpha').$$
These densities are originally defined for $\alpha>0$, but the equations allow
us to define $\rho_k(-\alpha)\equiv\rho_k(\alpha)$ in order to rewrite the
integrals to go from $-\infty$ to $\infty$. The kernels in these equations are
defined most easily in terms of their Fourier transforms:
\eqn\forphi{\tilde \phi_{kl}(x)=
\int^{\infty}_{-\infty} {d\alpha\over 2\pi} e^{i\alpha tx/\pi }
\phi_{kl}(\alpha)=
\delta_{ab}-2{\cosh x \sinh(t-k)x
\sinh lx \over \sinh  x\sinh tx},}
for $k\geq l$ with $\phi _{lk}=\phi_{kl}$, and
\eqn\forps{\eqalign{\tilde a_k&={\sinh (t-k) x
\over\sinh  tx}\cr
\tilde u_k&=2 {\sinh (t-H)x \sinh kx  \over \sinh x \sinh tx} +
{\sinh (t-2k)x/2\over \sinh  tx/2} - 1\cr
\tilde u_a&=2{\sinh Hx \over \sinh  tx}- {\sinh
(t-2)x/2\over \sinh tx/2} - 1.\cr}}

The boundary manifests itself in the first term in $u_k$; notice that even for
$h=0$ it still modifies the equations. A few technicalities account for the
other
terms (these are  relevant here because we are interested in subleading
boundary effects). The second term in $u_k$ arises from the fact that the sum
in \foral\ does not include the term $m=j$; the integration over densities
includes such a contribution and so it must be subtracted off by hand. The
third term in $u_k$ arises because $\rho$ and $\rho^h$ are defined for allowed
solutions, while as already explained, $\alpha=0$ is not allowed because it
does not give a valid wavefunction. Since it is a valid solution of \foral\ but
is not included in the densities, we must subtract an explicit ${2\pi\over
2N}\delta(\alpha)$
(corresponding to $1/2N$ in Fourier space).
One expects all other corrections to be suppressed by additional powers of
$1/N$.

\bigskip
\noindent{\bf 3. The physical Bethe ansatz equations}
\smallskip
We now derive the physical Bethe ansatz  equations by extending the computation
done in \KR\ for periodic boundary conditions. The two
cases $\epsilon=\pm 1$ are of  course very different.
\medskip
\noindent{\it Case $i$: $\epsilon=1$}
\smallskip
In this case the ground state is filled with anti-strings so that $\rho_k=0$
and $\rho_a^{h}=0$. We can use \foral\ to find the densities in the ground
state, but
we will not need these explicitly here. Excited states are given by
including string solutions ($\rho_k\ne 0$) and/or holes in the antistring
distribution ($\rho_a^{h}\ne 0$). The physical equations are obtained by
rewriting the bare equations \bba\ so that physical densities (i.e.\  $\rho_k$
and
$\rho_a^h$) appear on the right-hand side. This is done simply in Fourier space
by solving for the Fourier
transform $\tilde\rho_a$ in the last equation in \bba\ and substituting it into
the others. The result is
\eqn\pbai{\eqalign{2\pi (\rho_k + \rho_k^h)&= {1\over 2}
\left[A_k(\alpha+\Lambda)+A_k(\alpha-\Lambda)\right] +
\Phi_{k,t-1}*\rho^h_{a} +\sum_{l=1}^{t-1}
\Phi_{kl}*\rho_l +{1\over 2N}U_k\cr
2\pi(\rho_{a} +\rho_a^h)&=
2\pi(\rho_{t-1}+ \rho_{t-1}^h)+{1\over 2N}(U_a- U_{t-1})\cr}}
where the kernels this time are
\eqn\keri{\eqalign{&\tilde \Phi_{kl}=\delta_{kl}-2{\cosh x
\cosh(t-1-k)x
\sinh lx\over \cosh (t-1)x
\sinh x}\qquad k,l\ne t-1;\ k\ge l\cr
&\tilde \Phi_{t-1,k}=-{\cosh x
\sinh  kx \over \cosh (t-1)x
\sinh x}\qquad k\ne t-1\cr
&\tilde \Phi_{t-1,t-1}=-{\sinh  (t-2)x \over
2 \cosh (t-1)x \sinh x},\cr}}
where $\Phi_{lk}=\Phi_{kl}$, and.
\eqn\forqp{\eqalign{&\tilde A_k={\cosh (t-1-k)x
\over\cosh (t-1)x}\qquad\qquad k\ne t-1\cr
&\tilde A_{t-1}= {1\over 2 \cosh (t-1)x}\cr
&\tilde U_k= 2{\cosh(t-1-H) x
\sinh kx \over \cosh (t-1)x
\sinh x} - {\cosh x\sinh kx\over\cosh(t-1)x\sinh x}
\left({\sinh (t-2)x/2\over\sinh tx/2}+1\right)\cr
 &\qquad\qquad\qquad +{\sinh (t-2k)x/2\over\sinh tx/2}-1\qquad\qquad  k\ne
t-1\cr
&\tilde U_{t-1}={\sinh (2t-2-H)x \over
\cosh (t-1)x \sinh x}-
{\sinh (t-2)x/2 \cosh {t\over 2}x
 \over \cosh (t-1)x \sinh x } -{\sinh (t-2)x\over 2\cosh (t-1)x\sinh x}-1\cr
&\tilde U_a={\sinh Hx \over \cosh (t-1)x
\sinh x}-{\sinh (t-2)x/2 \cosh {t\over 2}x
 \over \cosh (t-1)x \sinh x } -{\sinh (t-2)x\over 2\cosh (t-1)x\sinh x}
-1.\cr}}
(We did not simplify some of the trigonometric sums to allow further
identification of the various terms.) These are the physical Bethe ansatz
equations, governing how the actual
quasiparticle excitations interact with each other. Notice that the kernels are
now symmetric, as opposed to \bba.
Each density corresponds to a quasiparticle; it is easy to read off the bulk
$S$ matrix elements.
In the next section, we will discuss how to do this and how to find the
boundary $S$ matrix.
\medskip
\noindent{\it Case $ii$: $\epsilon=-1$}
\smallskip
In this case the ground state is made of real solutions.
The physical densities are therefore $\rho_1^h,\rho_k$ and $\rho_a$.
 Eliminating $\rho_1$ from the right hand side of the bare equations gives
\eqn\neweqi{\eqalign{2\pi(\rho_1+\rho_1^h)&={1\over 2}[s(\alpha-\Lambda)+
s(\alpha+\Lambda)]+s*a_1^{(t-1)}*\rho_1^h\cr
&\qquad\qquad -\sum_{k=2}^{t-1}a_{k-1}^{(t-1)}*\rho_k+a_{t-2}^{(t-1)}*
\rho_a+{1\over 2N}u_0\cr
2\pi(\rho_k+\rho_k^h)&=a_{k-1}^{(t-1)}*\rho_1^h+\sum_{l=2}^{t-1}
\phi_{k-1,l-1}^{(t-1)}
*\rho_l-\phi_{k-1,t-1}^{(t-1)}*\rho_a+{1\over 2N}u_{k-1}^{(t-1,H-1)}\cr
2\pi(\rho_a+\rho_a^h)&=2\pi(\rho_{t-1}+\rho_{t-1}^h)+{1\over
2N}\left[u_a^{(t-1,H-1)}
-u_{t-2}^{(t-1,H-1)}\right],\cr}}
where the quantities $a^{(t-1)},\phi^{(t-1)}$ and $u_{k-1}^{(t-1, H-1)}$ are
the same as in the bare
equation \forphi-\forps\ with $t,k,H$ there replaced by $t-1,k-1,H-1$
respectively, and
$$\tilde u_0={\sinh (t-H)x\over \cosh x\sinh (t-1)x}+
{\cosh tx/2\sinh (t-2)x/2\over \cosh x\sinh (t-1)x}
+{\sinh (t-2)x\over 2\cosh x\sinh(t-1)x}-1$$
$$\tilde s={1\over 2 \cosh x}.$$
It is more difficult to read off the quasiparticle spectrum and the bulk and
boundary
$S$ matrices here because the bulk scattering is non-diagonal and most
densities correspond in fact to pseudoparticles. We will discuss this in the
next section.
\bigskip
\noindent{\bf 4. The $S$ Matrix}
\smallskip

In this section we derive the physical Bethe ansatz equations from the bulk and
boundary $S$ matrices by quantizing the momenta of a set of relativistic
quasiparticles on a line. We identify these with the physical equations derived
from the lattice model in the cases $\epsilon=\pm 1$.

In a relativistic quantum field theory the energy
and momentum of a particle of mass $m$ can be parametrized in terms of the
rapidity $\t$: $(E,P)=(m\cosh\t,m\sinh\t)$. Lorentz invariance requires that
$S$ matrix elements depend only on rapidity differences.

First let us discuss the case when the $S$ matrix is diagonal. When there are
$p$ different species of particle, we have the two-particle $S$-matrix elements
$S_{bc}(\t_1-\t_2)$, where $b$ and $c$ run from $1$ to $p$. These $S$-matrix
elements give the phase shift in the wavefunction when two particles are
exchanged. We also have the boundary $S$ matrix elements $R_b(\t)$, which gives
the phase shift when a particle of species $b$ bounces off a wall and changes
its rapidity from $\t$ to $-\t$. (We do not consider the case where a particle
changes species when bouncing off the boundary.)
We have a gas of ${\cal N}$ particles on a line of length $L$, with the $i$th
particle of species $b_i$. For the fixed boundary conditions we consider, we
demand that the wavefunction vanish at both ends of the line. This requires
making a stationary wave of states with opposite momenta,
with the momenta subject to the constraints
\eqn\quant{\eqalign{e^{iLm_{b_i}\sinh\t_i}&\prod_{j=1,j\ne i}^{\cal N}
S_{b_ib_j}(\t_i-\t_j)R_{b_i}(\t_i)\cr
&=e^{-iLm_{b_i}\sinh\t_i}\prod_{j=1,j\ne i}^{\cal N}
 S_{b_ib_j}(-\t_i-\t_j) R_{b_i}(-\t_i)\cr}}
for all $i$. When the scattering is trivial, this reduces to the familiar
$P=n\pi/L$, where $n$ is a positive integer. Unitarity requires that
$S^{-1}(\t)=S(-\t)$ and $R^{-1}(\t)=R(-\t)$, so this can be rewritten as
\eqn\quantii{e^{2im_{b_i}\sinh\t_iL}\prod_{j=1,j\ne i}^{\cal N}
S_{b_ib_j}(\t_i-\t_j)S_{b_ib_j}(\t_i+\t_j)R^2_{b_i}(\t_i)=1.}
Alternatively we can deduce these equations from systematic application of
 the Zam\-olod\-chi\-kov-Faddeev algebra together with its boundary
counterpart.

As in the lattice model, we define densities of solutions of \quantii\ so
that $(\rho_b(\t)+\rho_b^h(\t))2Ld\t$ gives the number of allowed rapidities
between $\t$ and $\t+d\t$ for species $b$, and $\rho_b$ gives the density of
filled states. Taking the logarithm of \quantii\ gives an equation of the form
\foral; taking the derivative of this gives an equation for the densities. We
again define the densities at negative rapidity by $\rho_b(-\t)=\rho_b(\t)$. We
have one equation for every type of particle:
\eqn\sba{2\pi(\rho_b+\rho_b^h)= m_b \cosh\t + \sum_{c=1}^p
\varphi_{bc}*\rho_c+ {1\over 2L}
\Theta_b,}
where
\eqn\forphi{\eqalign{\varphi_{bc}(\t)&=-i {d\over d\t}
\ln S_{bc}(\t)\cr
\Theta_{b}(\t)&=-i {d\over d\t} 2\ln R_{b}(\t) + i {d\over d\t} \ln S_{bb}(2\t)
-2\pi \delta(\t)\cr}}
The additional terms in $\Theta_b$ result from the same considerations as in
the
lattice model: the facts that the product in \quantii\ has no term $i=j$ and
that the solution with $\t=0$ should not be included. As before, we expect
additional corrections to be suppressed by powers of $1/L$.

The equations \sba\ are the physical Bethe ansatz equations derived directly
from the $S$ matrix description. They enable us to compare a diagonal $S$
matrix to lattice results. We do this in case {\it $i$} below. If the
scattering is not diagonal, the analysis is more complicated: one must
``diagonalize'' the multiparticle state, including the boundary $S$ matrices.
Basically, one must find the eigenvalues $\Lambda(\t_i|\{\t_j\})$ of the object
\eqn\transf{T(\t_i|\{\t_j\})_{bc}R_c(\t)
T^{-1}(-\t_i|\{\t_j\})_{cb}R^{-1}_{b}(-\t_i)}
where $T(\t_i|\{\t_j\})_{bc}$ is the matrix which describes scattering a
particle of species $b$ and rapidity $\t_i$ through all the other particles and
ending up with that particle being of species $c$.
The generalization of \quantii\ then is
\eqn\quantiii{e^{2im_{b_i}\sinh\t_iL} \Lambda(\t_i|\{\t_j\})=1.}
To diagonalize \transf, one uses the Bethe ansatz and the analysis proceeds as
before, but the calculation is a little delicate.
It is however simplified by using results of  \skly; we  discuss this in case
{\it $ii$} below.
\medskip
\noindent{\it Case $i$: $\beta^2=8\pi/t$}
\smallskip
As discussed in the introduction, the scaling  limit of the inhomogeneous
XXZ model with $\epsilon=1$ is the sine-Gordon model at $\beta^2=8\pi/t$.
The particles and their bulk $S$ matrix in this case are well known \ZandZ.
There are $t-2$ breather states with mass $m_a= 2 M \sinh{\pi a \over 2(t-1)}$,
along with a soliton and an antisoliton of mass $M$. All of these states
scatter diagonally with each other. The boundary $S$ matrices for the soliton
and antisoliton are conjectured in \GZ, and those for the breathers in \ghosi.

Using these $S$ matrix elements, we can now compare the physical Bethe ansatz
equations \sba\ with the lattice ones \pbai.
First we have to extract the scaling limit of the lattice model.
 This is easily done
by taking the limit $\Lambda\rightarrow\infty$ at finite bare rapidity
$\alpha$.
The source term in equations \pbai\ reproduces the expected spectrum with
$$M\propto {1\over\Delta}e^{-t\Lambda/2(t-1)}$$
and the physical rapidity
$$\theta={t\over 2(t-1)}\alpha,$$
provided we identify
the breathers with the string solutions $k=1\dots t-2$, the soliton
with the string solution $t-1$, and the antisoliton with the antistring $a$.

It is already known \KR\ that the kernels $\Phi_{jk}$ in \keri\
are the same as the $\varphi_{bc}$ that result from the known $S_{bc}$
scattering
 matrix.
It is straightforward but rather  tedious to check that the kernels
 $U_{k}$ in \forqp\ too are the same as
$\Theta_b$ resulting from \forphi\ and the $S_{bc}(\t)$ and $R_b(\t)$ of
\refs{\ZandZ,\GZ,\ghosi}, completing the boundary identification. Let us
discuss
 the case of soliton and antisoliton
 only for simplicity, denoting them $+$ and $-$ respectively. Using the
integral representation for the log of a
gamma function,  the result of \GZ\ can be written
$$\eqalign{-i\ln R_{\pm}(\theta)=\int &{dx\over x}
\sin{2\over\pi}(t-1)x\theta\ \cr
&\times\left[{\sinh (t-1\pm 2\xi/\pi)x
\over 2\cosh (t-1)x\sinh x} - {\sinh {3\over 2}(t-1)x\sinh ({t\over 2}-1)x
\over\sinh x/2\sinh 2(t-1)x}\right],\cr}$$
where $\xi$ depends on $\phi_0$. Comparing this with \forqp\ allows us to
identify
$$\xi={\pi \over 2}(t-1-H).$$
To identify the remaining terms we recall that
$$-i\ln S_{++}(\theta)=-\int {dx\over x}
{\sinh (t-2)x\over 2\sinh x\cosh (t-1)x}\sin {2\over\pi}(t-1)x\theta.$$
The identity
\eqn\trigidt{
\eqalign{{\sinh 3(t-1)x/2\sinh (t-2)x/2\over\sinh x/2\sinh 2(t-1)x}
-{\sinh (t-2)x/2\cosh tx/2
\over 2\cosh (t-1)x\sinh x}=\cr
{\sinh (t-2)x/2\over 4\sinh x/2\cosh (t-1)x/2}+
{\sinh (t-2)x\over 4\sinh x\cosh (t-1)x}\cr}}
completes the desired identification.

This confirms the $S$ matrices conjectured in \refs{\GZ,\ghosi}
 at these values of $\beta$. We emphasize  that it is absolutely crucial to
have included the correction terms to $u_k$ in \forps\ and $\Theta_b$ in
\forphi\ to get this result.

\medskip
\noindent{\it Case $ii$: $\beta^2=8\pi(t-1)/t$}
\smallskip
The continuum limit of the lattice model at $\g=\pi/t$ is the sine-Gordon model
at $\beta^2=8\pi(t-1)/t$. The only states in the spectrum are the soliton and
antisoliton, and they scatter with a nondiagonal $S$ matrix \ZandZ. With our
fixed boundary conditions, soliton number is conserved, so the scattering off
the wall is diagonal but only if $h=0$ ($H=t-1$) is it equal for soliton
and antisoliton \GZ.

To derive the Bethe ansatz equations   for this nondiagonal $S$ matrix, we need
the
eigenvalues of the matrix \transf.  We now discuss  some technical
aspects of this computation;
the uninterested reader can jump directly to the next paragraph. The equivalent
problem for periodic boundary conditions, where one diagonalizes $T_{bb}$, is
well understood. In this case, a simplifying feature
is that $S(0)$ is simply the permutation
operator. Hence, instead of $T_{bb}$ one can
consider the matrix describing the scattering of one particle through all the
others including itself. The unwanted contribution
of a particle scattering with itself is trivial anyway and does not have to be
discarded. The advantage is that the new diagonalization problem fits perfectly
in the inverse scattering framework and is easy to solve. In the presence of a
boundary the
situation is different.  If one tries the same approach there are now two
unwanted contributions, since the particle
at rapidity $\theta$ can ``scatter with itself'' at rapidity $\pm\theta$. The
process with $S(\t--\t)=S(2\t)$ leads to non-trivial terms. When the scattering
was
diagonal this was easy to discard by hand, but the situation is more
complicated here.  On the other hand, the inverse scattering formalism
in the presence of boundaries developed e.g.\ in \refs{\skly,\boundrefs}
 is treating a slightly different problem:
 $R^{-1}_b(-\t)$ is replaced there  with $R^{-1}_b(-\t-i\pi)$.
Fortunately these two complications essentially cancel each other;
changing the rapidity $-\t$ to $-\t-i\pi$
acts as a projector which cancels out the unwanted processes resulting from the
self-scattering term. The final  expression for the eigenvalue
$\Lambda(\t_i|\{\t_j\})$ turns out to be the same as that of
\skly, up to some missing  prefactors  that produce a phase, as required.
The complete proof is rather tedious and will be provided upon request. The
interested reader can also check the final result by direct calculation
in the two-particle case.

The result is that the auxiliary problem determining the eigenvalue $\Lambda$
is similar to the bare problem, up to a redefinition of some parameters.
Let us  define
$\sigma(\t)$, which is the density of particles (soliton or antisoliton).
In this case we have
$$-i\ln R_{\pm}(\theta)=\int {dx\over  x}\left[{\sinh
(1\pm 2\xi(t-1)/\pi)x
\over 2\cosh x\sinh (t-1)x}+
 {\sinh {3\over 2}x\sinh
(t-2)x/2\over\sinh (t-1)x/2\sinh 2x}\right]\sin {2\over\pi}x\theta$$
together with
$$-i\ln S_{++}(\theta)=
-\int {dx\over x}{\sinh (t-2)x\over 2\cosh x\sinh (t-1)x}
\sin {2\over\pi}x\theta.$$
The first part of the quantization equation
(the derivative of the log of \quantii) reads
\eqn\sysi{2\pi(\sigma+\sigma^h)=m\cosh\theta-{i\over 2L}{d\over d\theta}
\ln\Lambda,}
with
$$-{i\over 2L}{d\over d\theta}\ln\Lambda=\left(-i{d\over d\theta}\ln
S_{++}(\theta)\right)*\sigma-\sum_{b=1}
^{t-2}a_b^{(t-1)}*\sigma_b+a_{t-2}^{(t-1)}*\sigma_a+{1\over 2L}U,$$
and
$$U=-i{d\over d\t}2\ln R_{+}(\t)+
i{d\over d\t}\ln S_{++}(2\t)-2\pi\delta(\t).$$
For the other densities $\sigma_b$ and $\sigma_b^h$ ($b=1\dots t-2$),
we have the bare equations \bba\ with three changes. First, $t$ is replaced by
$t-1$ because there are $t-1$ solutions to the new Bethe equations. There is
also a new source
term $a_b^{(t-1)}*\sigma$ because the ``vertices'' in the problem of
diagonalizing \transf\
correspond to the real particles, and have fluctuating numbers and rapidities.
These two changes are identical to the case with periodic boundary conditions.
In addition, we have a boundary term which from \skly\ is the same as the
one in the bare equations with $H$ replaced by $H-1$. This shift is of course
related to the shift of $t$. For instance, if in the original bare equations
$h=0$ so that $H=t-1$, then in the auxiliary problem we again expect
soliton-antisoliton symmetry yielding $H=(t-1)-1$.

We can thus identify \sysi\ and the equations for $\sigma_b$ with \neweqi\
provided we set
$$\xi={\pi\over 2}\left(1-{H\over t-1}\right)$$
by using again the identity \trigidt\ with $x$ replaced by $(t-1)x$ and $t-1$
by $1/(t-1)$. This confirms results of \refs{\GZ,\ghosi}\ at values of $\beta$
where the scattering is non-diagonal.
\bigskip
\noindent{\bf 5. Conclusion}
\smallskip

We have verified in both attractive and repulsive regimes that the boundary $S$
matrices conjectured in \refs{\GZ,\ghosi} are in fact correct. We think that
the main interest of our computation lies in the delicate analysis of all the
boundary terms, which are crucial to make the computation consistent. In
particular, observe from the example of $\alpha=0$ that at order $1/N$, the
modification of even a single root in the Bethe equations affects the results.
This puts the string hypothesis to a much stronger test than the bulk
computations done so far. If we believe that the picture of \GZ\ is consistent
and that the inhomogeneous XXZ model provides also a regularization of the
sine-Gordon model in the presence of boundaries, we are forced to conclude that
strings are the {\bf only} solutions of the Bethe equations that contribute to
the scaling limit, i.e.\ to the low-energy excitations. This does not
contradict the known examples of non-string solutions, which either exist in
finite numbers \refs{\DL,\nonstrings}\
(so we show that the finite number is zero in the scaling limit) or are at very
high energy \EKS.

The next step is to study the system at finite temperature and to derive the
thermodynamic Bethe ansatz equations for the free energy. This should yield the
``$g$-function'', which measures the number of degrees of freedom on the
boundary \AL. At a fixed point, this can be calculated from conformal field
theory \refs{\CarVer,\AL}, and was calculated for the free boson (the critical
limit of the sine-Gordon model) in \EA.  It is simple to get the bulk part of
the free energy from the physical Bethe ansatz equations, but additional
subtleties appear because the $g$-function is a subleading term. For example,
the usual thermodynamic Bethe ansatz free energy is the log of the partition
function only at leading order in
$L$; there are system-dependent corrections at lower order. However, when one
looks at massless $S$ matrices to describe a flow between ultraviolet and
infrared fixed points, the ratio $g_{UV}/g_{IR}$ should be independent of these
corrections. It is indeed simple to calculate this ratio from the $S$ matrix,
for example, in the Kondo problem \refs{\AFL,\pkondo} or in the flows between
the minimal models. We hope to discuss these more physical aspects in the near
future.

\bigskip

{\bf Acknowledgments}: This work was supported by the Packard Foundation, the
National Young Investigator program (NSF-PHY-9357207) and  the DOE
(DE-FG03-84ER40168). We thank A. Delfino and G. Mussardo for very useful
discussions on boundary problems.

\listrefs
\end